\documentclass[traditabstract]{aa}
\usepackage{graphicx,psfrag,amsmath,color}

\usepackage{ulem}
  
\usepackage{amsmath}

\usepackage{graphicx}
\usepackage[varg]{txfonts}
\usepackage{natbib} 
\bibpunct{(}{)}{;}{a}{}{,} 
\definecolor{violet}{rgb}{0.5,0,0.5}

\begin{document}

\title{Collisionless Boltzmann  equation approach for the  study \\of  stellar discs within barred galaxies}

\titlerunning{CBE approach for the  study of  stellar discs within barred galaxies}

\author{O. Bienaym\'e\inst{1}, }

\institute{Observatoire astronomique de Strasbourg, Universit\'e de Strasbourg, CNRS,  11 rue de l'Universit\'e, F-67000 Strasbourg, France 
 }


\date{Received / Accepted}

\abstract{
We have studied the kinematics of  stellar disc populations within the solar neighbourhood in order to find the imprints of the Galactic bar. We carried out the analysis by developing  a numerical resolution of the 2D2V (two-dimensional in the physical space, 2D, and two-dimensional in the velocity motion, 2V) collisionless Boltzmann equation (CBE) and modelling the stellar motions within the plane of the Galaxy within the solar neighbourhood. We recover similar results to those obtained by other authors using N-body simulations, but we are also able to numerically identify faint structures thanks to the cancelling of the Poisson noise. We find that the ratio of the bar pattern speed to the local circular frequency is in the range $\Omega_B/\Omega$ = 1.77 to 1.91.   If the Galactic bar angle  orientation is within the range from 24 to 45 degrees, the bar pattern speed is between 46 and 49 km.s$^{-1}$.kpc$^{-1}$.
}


\maketitle

\section{Introduction}
Since the analysis of the Hipparcos observations, it has been known that the velocity field of stars within the solar neighbourhood is highly structured. The wavelet analysis of the velocity distribution of stars has shown that structures are present at all scales \citep{che98, che99} with the smallest scales being of the order of 3 km.s$^{-1}$. Many of these structures correspond to  identified moving groups or streams. At the largest scale, the Hercules stream is a distinct structure clearly recognisable with Hipparcos observations \citep{deh98}. Later, these results were confirmed and improved with the availability of radial velocities in complement to Hipparcos proper motions and parallaxes \citep{fam05,ant08}. It is now generally admitted that most of the  moving groups  does not result  from the dissolution of stellar clusters. The reason being that  stars from  a moving group are generally  not coeval and have different chemical abundances  \citep[see][]{fam08, pom11,ben14}  \citep[but see][]{tab17}.  It is now accepted that most moving groups  must result from a dynamical process. The most popular explanation concerning the Hercules streams is  the proximity of the Sun  to the OLR (outer Lindblad resonance) produced by the rotating Galactic bar. Other dynamical explanations were also proposed as streams induced by chaotic motion \citep{fux01,rab98} or by the ILR of spiral arms \citep{sel10}. These explanations hold for the two main structures within the ($V_R, V_\theta$) velocity field. It is suspected that non stationarity, combined to the effects of the bar and spiral arms, leads to the  substructures seen at the smallest velocity scales.  The recent surveys RAVE \citep{kun17}, SEGUE \citep{yan09}, LAMOST \citep{xia17}, and now Gaia DR1 \citep{gai16} have allowed us to probe larger distances from the immediate solar neighbourhood and to compare the  structures of the velocity field  at different positions within the Galaxy \citep{ ant15,mon16a,mon16b}, and  above the Galactic mid-plane    \citep{mon14}.

The existing models of  dynamical processes that  explain the observed moving groups are based on  different approaches:  firstly, analytical study  of local resonances \citep{mon16a}, secondly, particle-test in a time evolving gravitational potential \citep{min07,min10,min11} or finally, live N-body simulations \citep{qui11}.  Realistic simulations based on N-body (particle test or live) are still limited by the number of particles and they poorly   model the faintest structures  that remain dominated by the Poisson noise. On the other hand, analytical models of resonances allow  to consider more precisely the presence of dynamical substructures within the velocity field, but this implies specific development for each resonance.

Here, to circumvent these  limitations,  we have considered   the resolution of the collisionless Boltzmann equation (CBE) that can be seen as an ideal N-body simulation in the limit of N tends  to the  infinity. This drastically different approach can be compared with other models. A significant  advantage   is   the easiest identification of the smallest substructures, not dimmed and confused  with the Poisson noise fluctuations. However some specific limitations related to the numerical resolution of the CBE  exists and will be discussed. Here, we have limited our study to a model that is two-dimensional in space and in velocities (2D2V)  of a galactic  stellar disc  within a rotating barred potential.

The numerical resolution of the CBE had not  been frequently envisaged in the context of galactic dynamics. It  does not allow to easily follow individual stars and orbits, and  it also requires large  numerical resources  of CPU and memory. Only recently have full 3D3V galactic evolution models been achieved \citep{yos13, sou16}. We mention some previous works based however on smaller dimensionality by \citet{col15} with a detailed bibliography, \citet{ala05} and also pioneering works to study the stability of galactic discs by  \citet{nis81} and \cite{wat81}.       We also  refer the reader to the 1D2V Vlasov-Poisson resolution by \citet{val05} \citep[see also][]{man02}. Our numerical resolution  of the CBE is  based on their numerical analysis.

To close this introduction, we note that the Boltzmann collisionless equation (so named on the recommendation of H\'enon),  is  frequently named the Vlasov equation although this is not historically justified \citep[see][for  a discussion]{hen82} while  other designations exist.

The paper proceeds as follows. Section 2 describes the numerical scheme to solve the CBE and Section 3 the galactic modelling (potential and initial distribution function of stars) with the results of some tests. Section 4 describes the results of Galactic modelling of stellar streams and gives the comparison with previous numerical studies of stellar streams followed by a conclusion (Sect. 5).

\section{Numerical integration of the CBE}

The numerical integration used to solve the CBE
 is based on the splitting method,  coupled with a finite difference upwind scheme. The algorithm is  borrowed from the works of  \citet{man02} and \citet{val05}
 that solve  the CBE in a  1D2V case (cartesian in coordinate and cylindrical in velocities). Their scheme provides a second order accuracy in space and time and  they give the  detail of the equations that they have developed. They are  succinctly reproduced here and adapted to the  2D2V case.

With  $f(x,y,x_x,v_y,t)$ the distribution function and $F_x, F_y$ the forces,  our CBE resolution is performed in cartesian coordinates:

$$
\frac{\partial f}{\partial t}
+v_x\frac{\partial f}{\partial x}  
+v_y \frac{\partial f}{\partial y}
+F_x \frac{\partial f}{\partial v_x}
+F_y \frac{\partial f}{\partial v_y}=0 \, .
$$

The two main steps  are, first, the time discretisation and, second,  the space and velocity discretisation. The splitting scheme allows us to split the evolution of the distribution function into two steps, one in the physical space, the other in the velocity space \citep{che76}.

The numerical scheme to solve the 2D2V Boltzmann equation can be written with the operators $\Lambda$ and $\Theta$, following a similar evolution operator notation used by \citet{man02}:

$$\Lambda_x=-v_x \frac{\partial}{\partial x}, 
\Lambda_y=-v_y \frac{\partial}{\partial y},
$$
$$
e^{\Lambda_{xy} \,\tau}=\exp\left(\, \frac{\Lambda_x \tau}{2}\,\right) 
\exp\left(\, \Lambda_y \tau\,\right) 
\exp\left(\, \frac{\Lambda_x \tau}{2}\,\right) $$
$$ f(x,y,\tau)= \exp \left( \Lambda_{xy} \tau \right) f(x,y,0)$$
$$\Theta_x=-F_x \frac{\partial}{\partial v_x},
\Theta_y=-F_y \frac{\partial}{\partial v_y},
$$
$$
e^{\Theta_{xy} \,\tau}=\exp\left(\, \frac{\Theta_x \tau}{2}\,\right) 
\exp\left(\, \Theta_y \tau\,\right) 
\exp\left(\, \frac{\Theta_x \tau}{2}\,\right) $$

$$ f(v_x,v_y,\tau)=\exp \left( \Theta_{xy} \tau  \right) f(v_x,v_y,0)$$
$$f(x,y,v_x,v_y,\tau)= e^{\Lambda_{xy} \,\tau/2}
e^{\Theta_{xy} \,\tau}
e^{\Lambda_{xy} \,\tau/2}
f(x,y,v_x,v_y,0)$$

with the forces  calculated just after the first half-time step.

Thus, the multidimensional CBE equation is splitted into four different one-dimensional equations of the general form

\begin{equation}
\label{eqAdvect}
\frac{\partial f}{\partial t} + \nabla . (Af) =0
\end{equation}

that are solved by   the van Leer's upwind scheme \citep{lee76,har82} described in \citet{val05}. This scheme  is second order in space and time and is stable under the Courant condition.

\paragraph{Test:}

Detailed tests are given in \cite{val05} in a  1D2V case.
Here, we want to model  distribution functions that are  stationary solutions of the BCE and we check that these solutions are not modified by numerical diffusion or oscillations. Since we want to model the stellar kinematics during a few dozen of rotation of a Galactic bar, we only need to check the stationarity over a few hundred dynamical times.  We determine the smallest size of  gaussian structures in the phase space that are not altered  by the numerical diffusion during that period of time. For the resolution of Eq.~\ref{eqAdvect},  we have also tested otsher schemes that are described and accuracy are detailed  in \cite{tor09}: a basic Godunov scheme, another      scheme of Godunov  with  a central difference scheme that is second-order, a Lax-Wendroff scheme, a Warming-Beam scheme (a second-order upwind method), and a Fromm scheme.

We have examined the impact of strong gradients and the numerical diffusion in    1D, 2D, and 3D cases, looking at the evolution of a distribution function gaussian in space and velocities. We have  checked the conservation of the initial distribution function with time and through  space using two types of models: the first have no forces but do have periodic boundaries in order to  examine a uniform advection (with 1, 2 or 3D).  A second series of test models have harmonic potential (2D), or using the following equation as a 3D case
$$ \frac{\partial f}{\partial t} + 
(y-z) \, {\rm d}f/ {\rm d}x\, +
(z-x) \, {\rm d}f/ {\rm d}y\, +
(x-y) \, {\rm d}f/ {\rm d}z\, 
=0$$
For this second series of models, the distribution function (DF) $f$ rotates around the origin (2D case) or around the axis of direction (1,1,1) (3D case). All these simulations show conditions under which   the shape of the DF remains unmodified.

From the different tested schemes, we have selected the van Leer scheme that is the most conservative,  the other ones being  more diffusive. In this case, a gaussian DF (in space and in velocity) remains unaffected as long as its width  is larger or equal to four pixels of the grid.  It does not vary by more than a few percent in amplitude for at least 300 dynamical times. 

More  accurate algorithms exist as for instance the positive flux conservation (PFC) scheme of  \citet{fil01} used by \citep{yos13}.  With this algorithm,  steeper gradients can be accurately modelled.  However,  we implement the Mangeney algorithm owing  to its relative simplicity to code up and since the $\sim 1\%$ precision reached is sufficient for our purposes.

\section{Galactic modelling}

We are interested in  modelling the velocity field of stars 
and in examining the impact of the galactic bar rotation within the solar neighbourhood.
For that purpose, we model the gravitational potential of the galaxy with an analytic axisymmetric disc that has a circular velocity curve rising from  the Galactic centre and becoming flat at large distances. The circular velocity curve is  given by:
$$v_c(R)=\frac{R}{\sqrt{a^2+R^2}} \,\,v_{\infty} \, .$$
We set $v_{\infty}=1$ and $a=0.3$. The analytic bar potential is modelled by including a bisymmetric perturbation to the axisymmetric potential  given by:
$$\Phi_b= \frac{\epsilon \,v_\infty^2}{2} \frac{R^3}{(a^2+R^2)} \exp(-R/R_b)  \cos\left(2 (\theta-\Omega_b t)\right)\, ,
$$
where $\epsilon$ gives the strength of the bar, and $R_b$ its radial extension. We set $R_b$=0.408. Then the ratio of the maximal tangential force $F_{bar,\theta}$  to the radial force of the disc is:
 
$$Q_\theta=F_{bar,\theta}/F_{disc}=\epsilon \, R \, \exp(-R/R_b)\, ,$$

\begin{figure}[!htbp]
\begin{center}\
\resizebox{8.5cm}{!}{\rotatebox{0}{\includegraphics{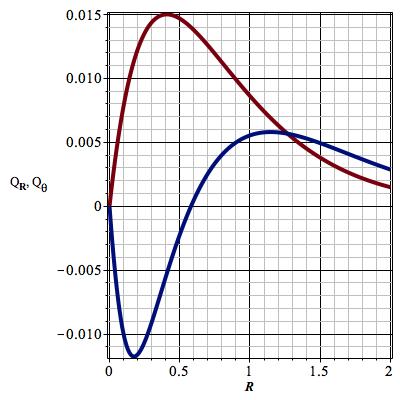}}}
\end{center}
  \caption{  Ratio of the tangential force, $Q_\theta$ (red), and radial force, $Q_r$ (blue), of the bar to the radial force of the disc, with $\epsilon$ = 0.1 and $R_b$=0.408, depending on the Galactic radius. }   
   \label{fig_ratio}
\end{figure}

Figure \ref{fig_ratio} shows $Q_R$ and  $Q_\theta$ for $\epsilon$ = 0.1,  the ratios of the  maximal radial and tangential bar forces to the radial force of the axisymmetric disc component. This can be compared with similar  figures of bar forces used for the determination of stellar galactic orbits \citep{ath83}.  We fixed the angular velocity  of the bar   $\Omega_p=1.6652$, so the Outer Linblad Resonance (OLR)  is exactly $R_{olr}=1$, the corotation radius  $R_{cor}$ = 0.52 and there is no ILR. In the case $\epsilon$ = 0.1, the ratio of bar to axysimmetric forces is $Q_R$ = 0.0084 at $R_{olr}$ The mass of the bar is increased linearly with time from a null mass at time $T=0$ to its maximum value at  $T=30$  (approximately eight rotations of the bar) and then its mass is set constant.

The integration grid is cartesian. Its lower and upper bounds in $x$ and $y$ are equal to $\pm1.5$,   and  are  $\pm1.3$ in $v_x$ and $v_y$ velocity coordinates. The 4D grid size is $288^4$ pixel size, corresponding to steps of 0.01 for position and 0.009 for velocity (scaling for a galaxy with $R_{olr}\,=\,8.5$\,kpc and $V_c\,=\,220$\,km.s$^{-1}$,  gives steps of 85\,pc and 2\,km.s$^{-1}$). Since the effective resolution is four pixels to avoid numerical diffusion or  numerical oscillations,   the  effective resolutions are 0.05 in $x$-$y$ and 0.036 in $v_x$-$v_y$. (respectively 340\,pc and 8\,km.s$^{-1}$).

We chose at $T=0$ an initial Shu-type distribution function \citep{shu69}. This distribution function depends only on $E$ and $L_z$ and the corresponding stellar density is nearly radially exponential    \citep{bie99} with a scale length $R_\rho$ = 0.29 (assuming $R_{olr}\,=\,8.5\,$kpc gives  $R_\rho\,=\,2.5\,$kpc).

$$
f(x,y,v_x,v_y)=
\frac{2 \Omega}{\kappa}
\frac{\Sigma}{2 \pi \sigma^2}
\exp \left[- \frac{E-E_c(L_z)}{\sigma^2} \right]
$$
with
$$
\Sigma= \Sigma_0 \exp [ -R_c(L_z)/R_\rho]
$$

The initial radial velocity dispersion  $\sigma_R$ is set constant. Here, we only present  results for  the two cases  $\sigma_R$ = 0.1 and 0.2  corresponding to 22\,km.s$^{-1}$ and 44\,km.s$^{-1}$ if $V_c$=220\,km.s$^{-1}$.

\paragraph{Test with axisymmetric potentials:}

We have  tested and verified the reliability of our numerical integrations in the 2D2V case. First, we  have considered  simple axisymmetric potentials ($\epsilon$ = 0) for which Shu distribution functions are  exact stationary solutions of the Boltzmann equation. Using such  DFs as initial conditions, we  look at the numerical constancy of the DF during 32 rotations of the bar ($t = 120$).

With a   $288^4$ grid size and a sufficiently high initial velocity dispersions  ($\sigma_R\ge$  0.1) and for radii  not too close from the centre, $R > 0.15$, and away from the outer boundary, $R < 1.35$, we find that the initial Shu DF is nearly invariant over a long period of time. The density distribution remains exponential and conserves  the same scale length ($R_\rho$ = 0.29 from $T$ = 0 to 120), while  the  density decreases by less than 0.1\,\% for each $\Delta T=20$ step, at the exception of  outer region, $R$ > 1.35, where the density decreases due to the lost flow moving outside of the grid and not compensated by an equivalent inflow. The velocity distribution $f(V_R,V_\theta)$ also stays nearly unchanged, with the velocity dispersion ($\sigma_R \sim 0.1$ and $\sigma_\theta \sim 0.07$) changing by less than  0.1\,\% for each $\Delta T=20$ step. With this grid size, the initial velocity dispersion cannot be  much smaller values without introducing  diffusion.

Close to the centre of the grid and for $V_\theta$ values close to zero, the $v_\theta$ component of the initial Shu DF  has a very strong gradient.  There, large   numerical  diffusion and  oscillations appear. Within these inner parts of the simulation $R~<~0.15$ and after a relatively short time ($t \sim 20$ or 5 bar rotation) the DF reaches a stationary state. 

Our (lack of)  boundary conditions (BCs) at the physical border of the grid implies that any flow moving outside of the 4D grid is lost. It results that the areas in the corners of the $x$-$y$ square domain ($R>1.5$)  are rapidly emptied. On a longer timescale, the density at $R>1.35$ slowly decreases. To avoid the sweeping of the grid in the outer $x$-$y$ corners, we have tried another BC by fixing the DF along the $x$-$y$ border  to the initial Shu DF.  Comparing both BCs does not show visible differences for the velocity distribution when $R \le 1.4$.

For the simulations  with  barred potentials studied in the next paragraph,  both BCs  lead to  identical numerical results for the stellar   distribution function when $R$ is smaller than 1.4. As the bar is progressively introduced, the DF widens  within the bar  and quickly evolves towards a stationary state.  After $T=32$,  the mass of the bar is set constant and  the DF quickly stops to  evolve.  Thus,  from $T=32$ to $T=113$ and within the frame rotating with the bar, the relative change of the density   is of the order of one percent.


\begin{figure*}[!htbp]
\begin{center}\
\resizebox{6.0cm}{!}{\rotatebox{0}{\includegraphics{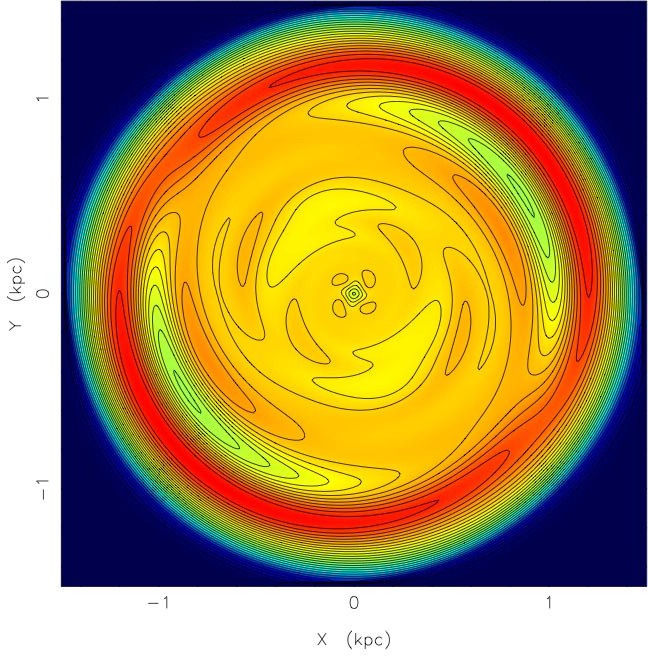}}}
\resizebox{6.0cm}{!}{\rotatebox{0}{\includegraphics{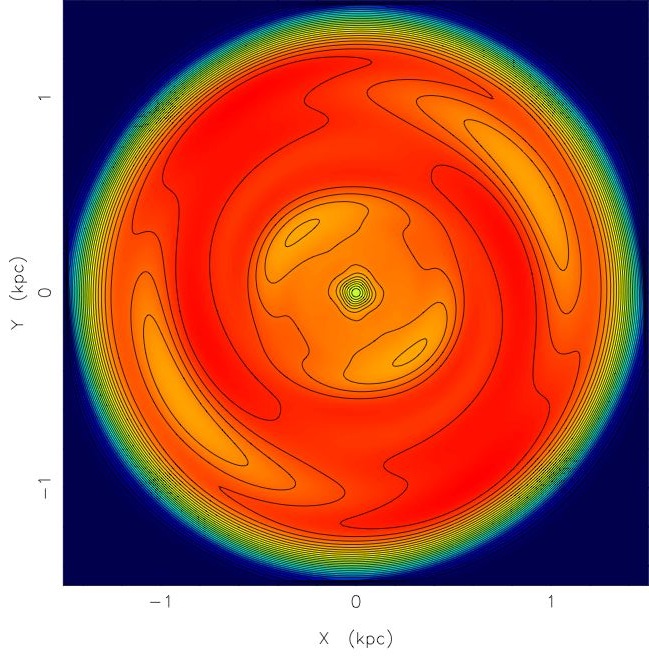}}}
\\
\resizebox{6.0cm}{!}{\rotatebox{0}{\includegraphics{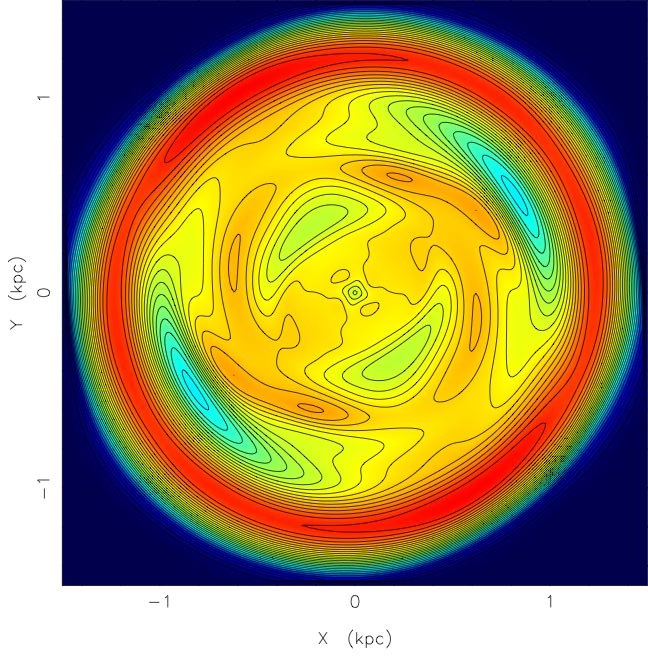}}}
\resizebox{6.0cm}{!}{\rotatebox{0}{\includegraphics{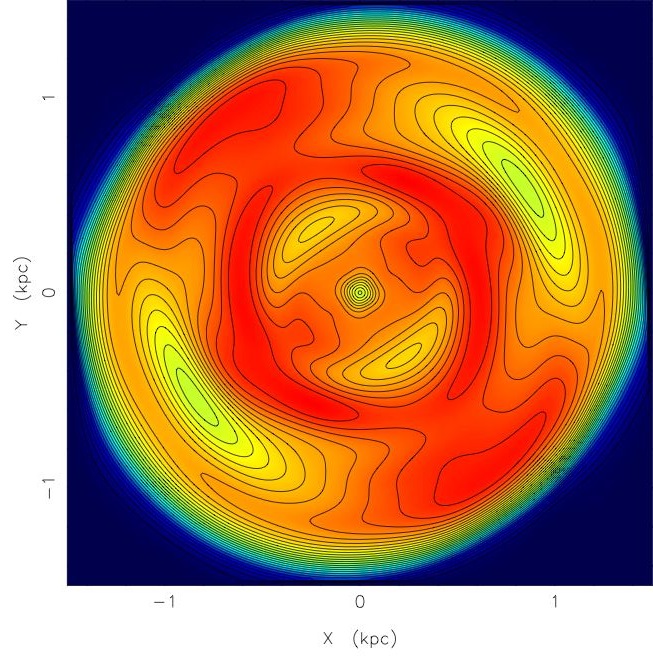}}}
\end{center}
  \caption{  
  Contrast of the stellar density distribution.
{\bf (upper left)} model A :   $\sigma_R=0.1$, $\epsilon=0.1$
{\bf (upper right)} model B :   $\sigma_R=0.2$, $\epsilon=0.1$
{\bf (lower left)} model C :   $\sigma_R=0.1$, $\epsilon=0.3$
{\bf (lower right)} model D :   $\sigma_R=0.2$, $\epsilon=0.3$.
The major bar axis  orientated at 34 degrees from the horizontal axis. Corotation at 
$R$ = 0.52 and OLR at $R$ = 1.     
Range of of colours is given with an arbitrary scale, different for each image to maximise the contrast (red is positive and blue negative).
  }
      \label{densABCD}
\end{figure*}

\section{Stellar steams}

The purpose of this work is the study of orbital resonances and stellar  streams within the Galactic disc. The identification of streams in the solar neighbourhood necessitates the statistical identification of overdensities in the velocity space. It implies the use of adapted statistical tools to minimize the number of false detections due the Poisson noise resulting from star counts \citep{che99}. Within the solar neighbourhood two main large streams, including the Hercules stream,  are identified. Smaller streams are also well identified. However, due to the limited number of Hipparcos stars within 150 pc distances and with accurate 3D velocities, the faintest visible structures may be   statistical noise. This difficulty is reduced by the recent advent of the Gaia DR1/TGAS survey \citep{gai16} and will be drastically minimized thanks to the forthcoming Gaia surveys that will have order of magnitudes larger samples of stars. 
Now the same difficulty arises analysing N-body simulations, especially analysing  a small space volume. Conversely, the CBE resolution allows us to cancel the Poisson noise and  to reach very high contrasts between the different streams. This is a significant advantage over N-body simulations. CBE resolution also allows us to compare entirely different methods to study the same physics.
 
\subsection{Models}

Here, we  have followed the evolution of the stellar disc distribution function within a barred potential.  We have  considered the impact of the two following parameters: the force of the bar, the velocity dispersion of the stellar component. In each case, we visually examined the distribution function. A wider variety of  models have been considered to explore other model parameters, but we only present  these  that mimic what we know of our Galaxy and of the stellar kinematics. 


\begin{figure*}[!htbp]
\begin{center}\
\resizebox{3.9cm}{!}{\rotatebox{0}{\includegraphics{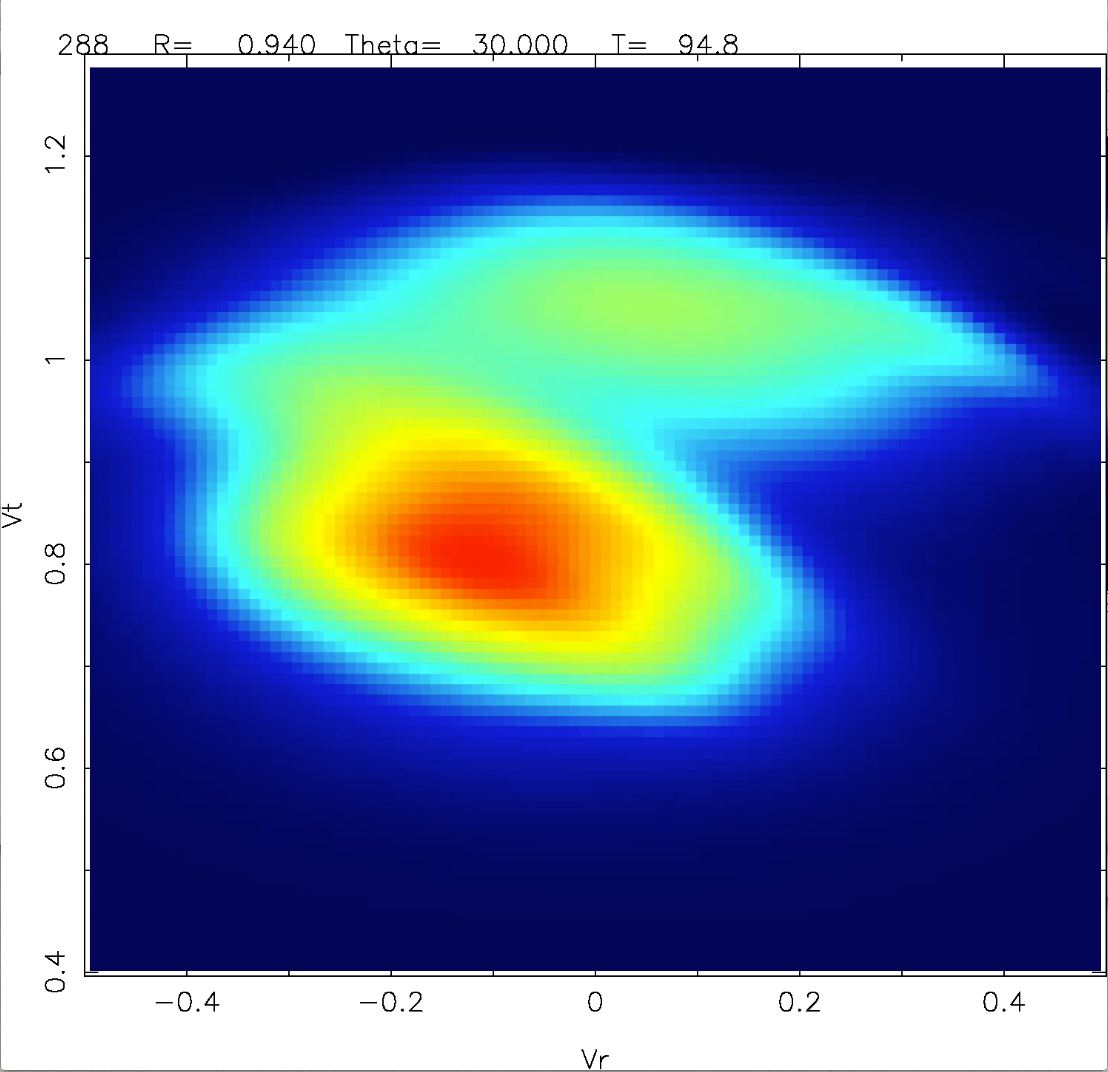}}}
\resizebox{3.9cm}{!}{\rotatebox{0}{\includegraphics{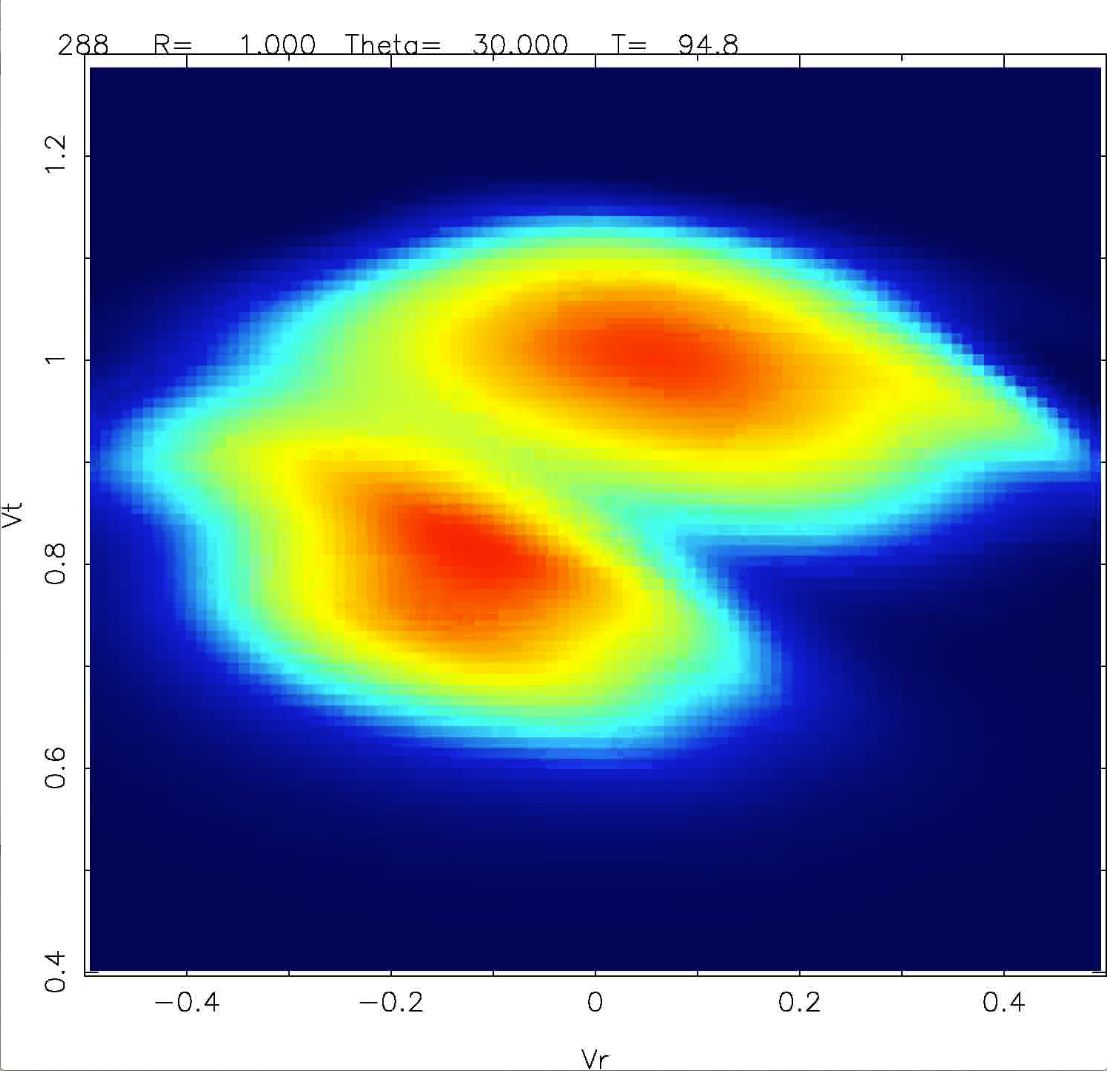}}}
\resizebox{3.9cm}{!}{\rotatebox{0}{\includegraphics{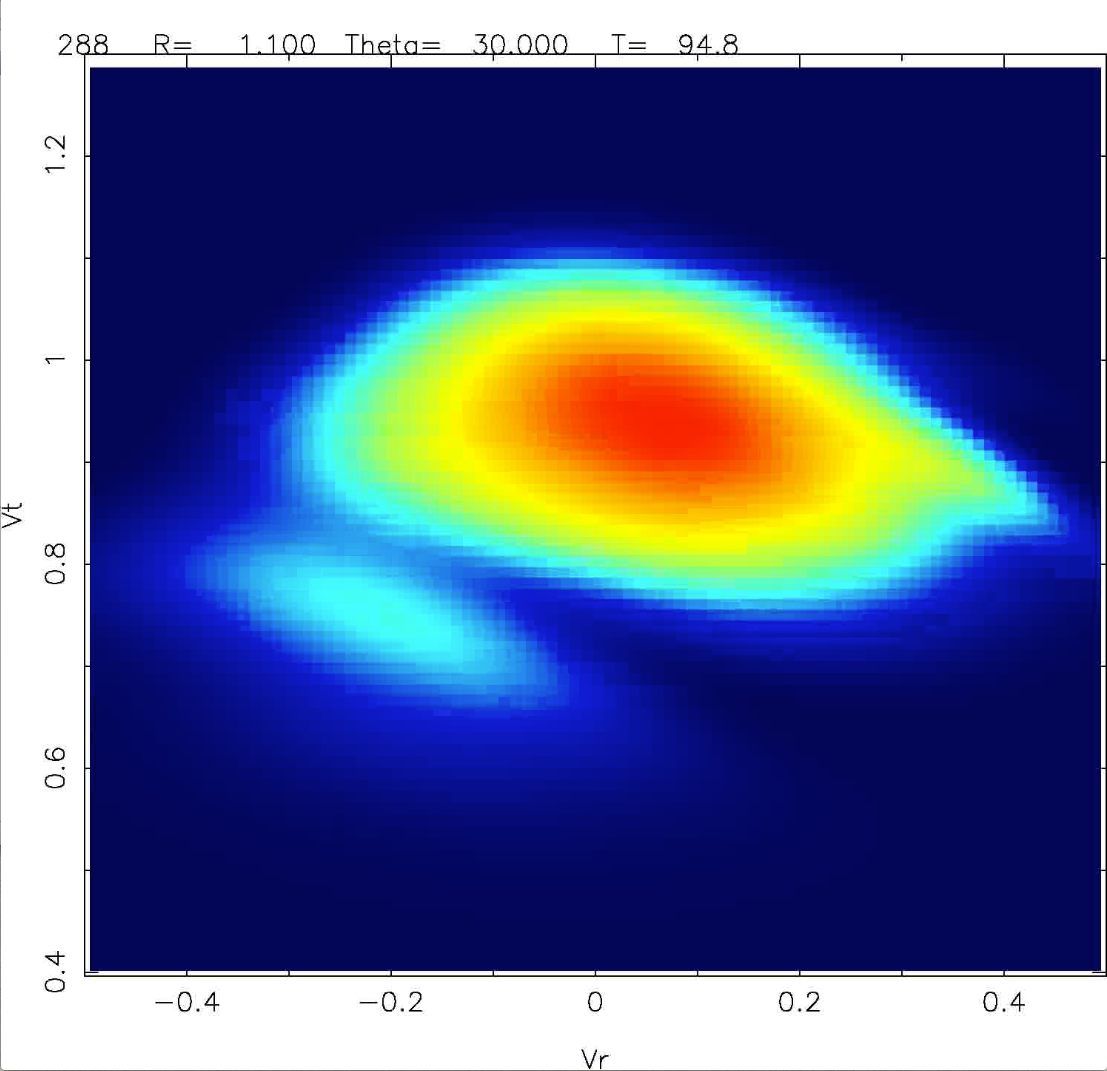}}}
\resizebox{3.9cm}{!}{\rotatebox{0}{\includegraphics{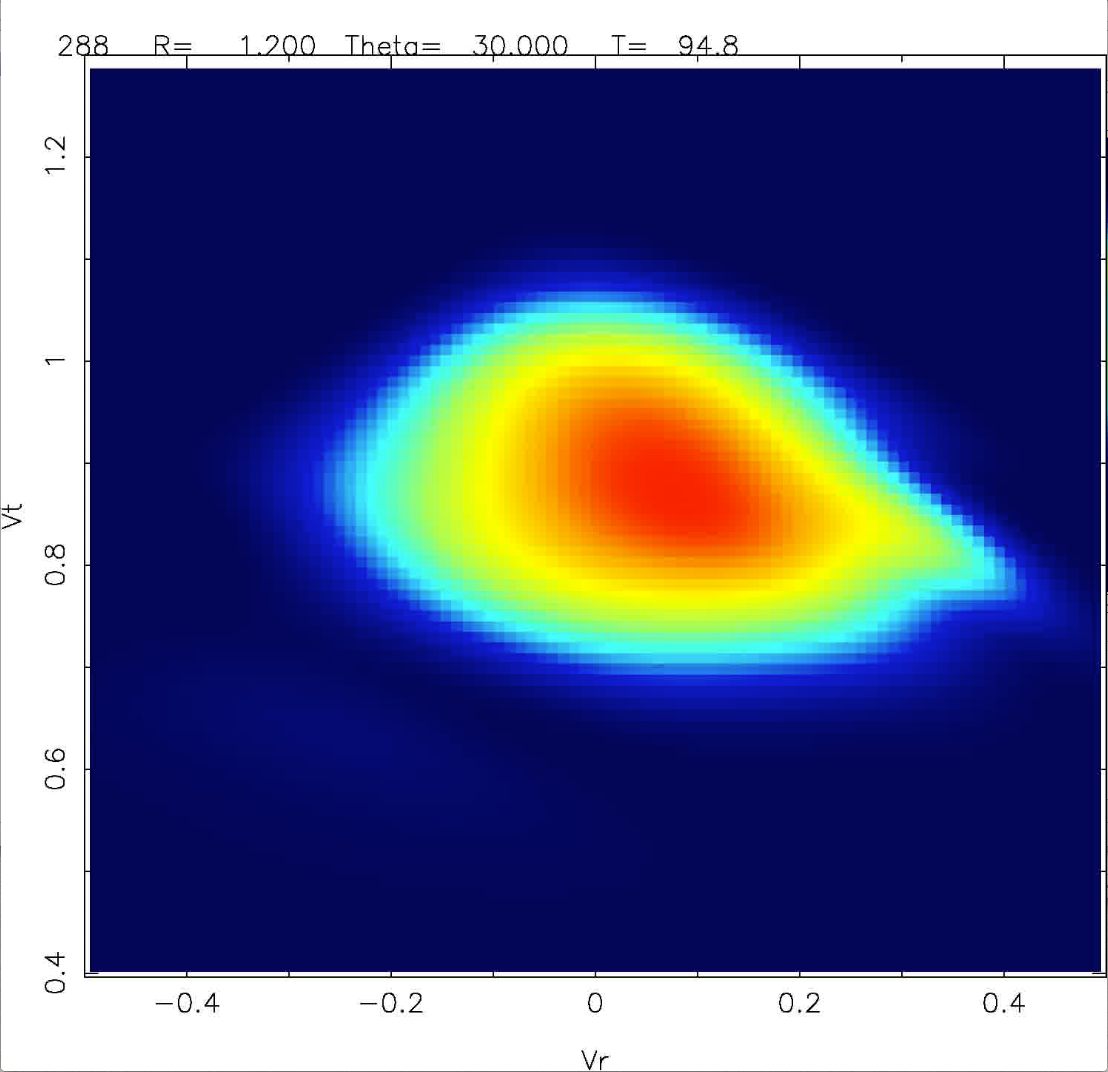}}}
\end{center}
  \caption{  ($V_R, V_\theta$) velocity distributions for model D at $R$= 0.94, 1.0, 1.1, 1.2  and $\theta=30$ degrees at $T$=94.8.}    \label{fdR}
\end{figure*}

\begin{figure*}[!htbp]
\begin{center}\
\resizebox{3.9 cm}{!}{\rotatebox{0}{\includegraphics{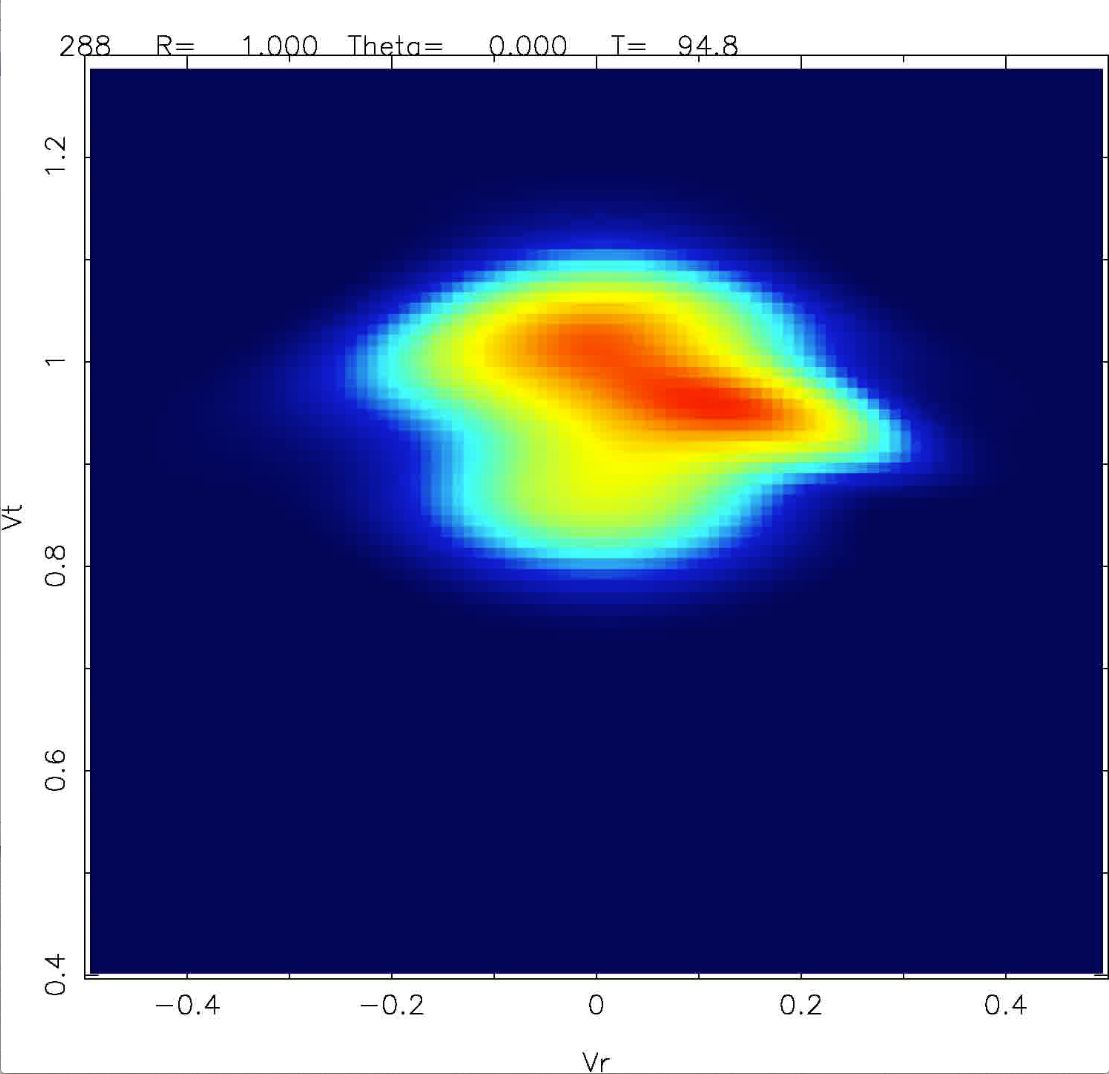}}}
\resizebox{3.9 cm}{!}{\rotatebox{0}{\includegraphics{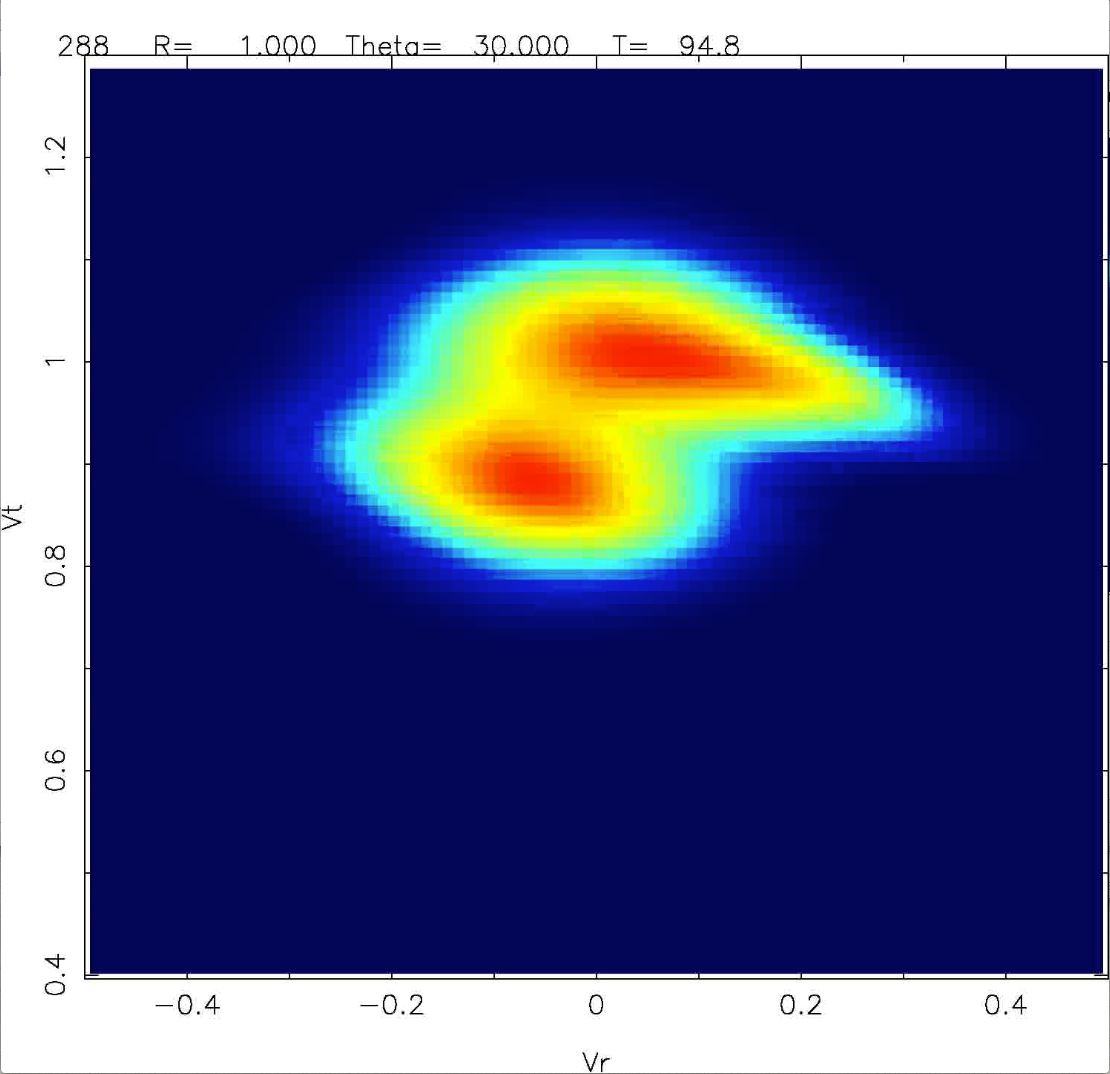}}}
\resizebox{3.9 cm}{!}{\rotatebox{0}{\includegraphics{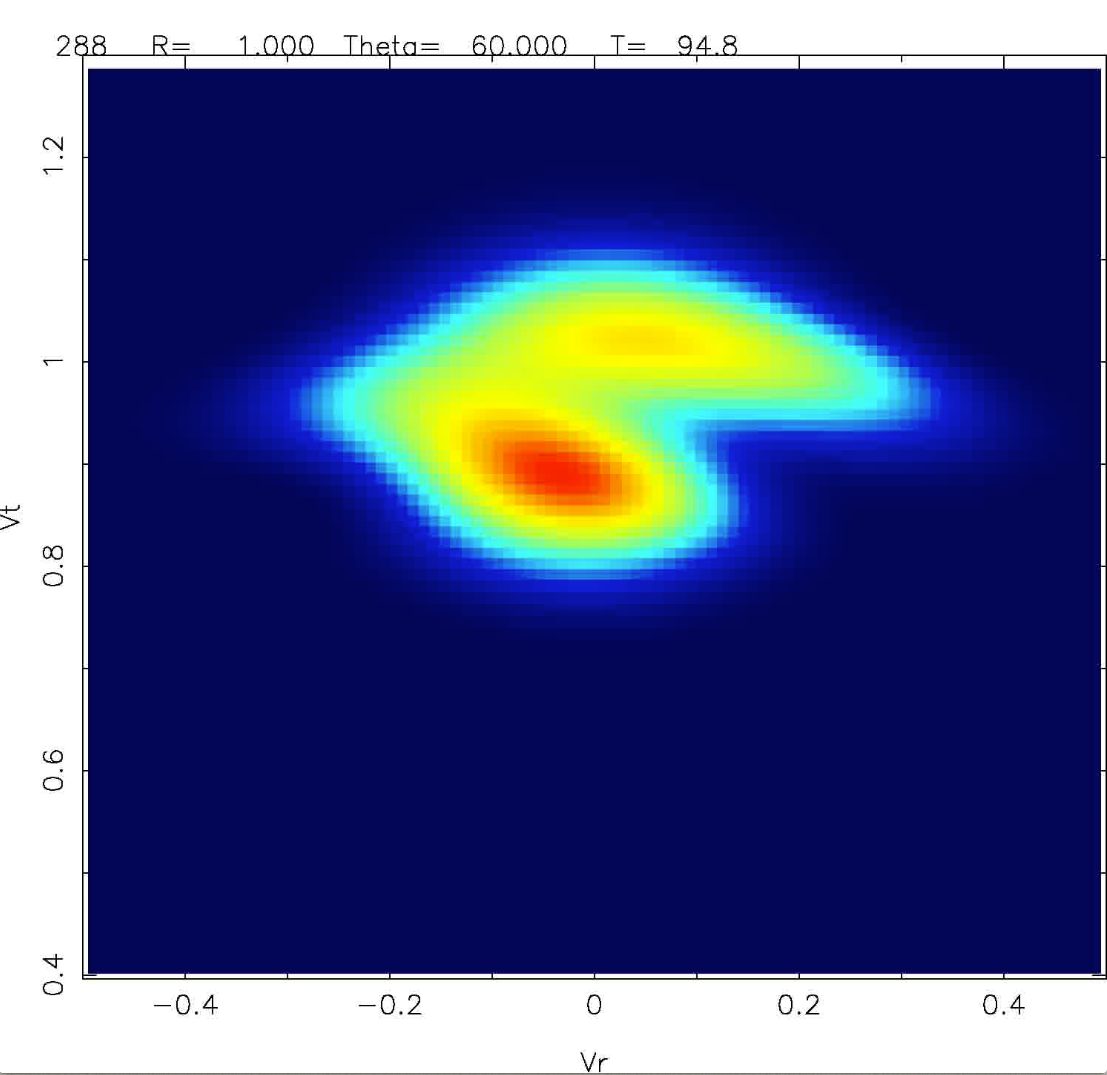}}}\\
\resizebox{3.9 cm}{!}{\rotatebox{0}{\includegraphics{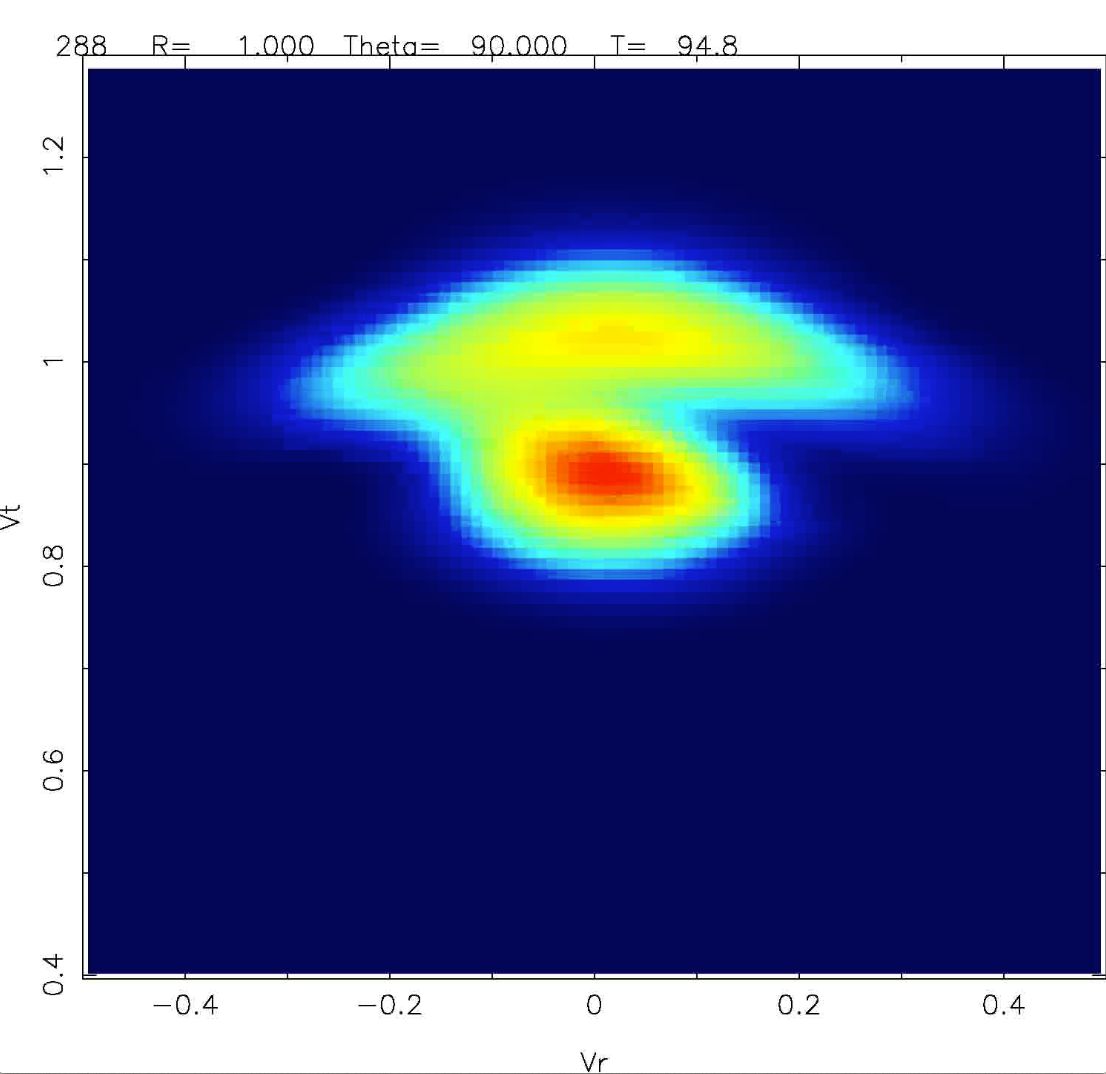}}}
\resizebox{3.9 cm}{!}{\rotatebox{0}{\includegraphics{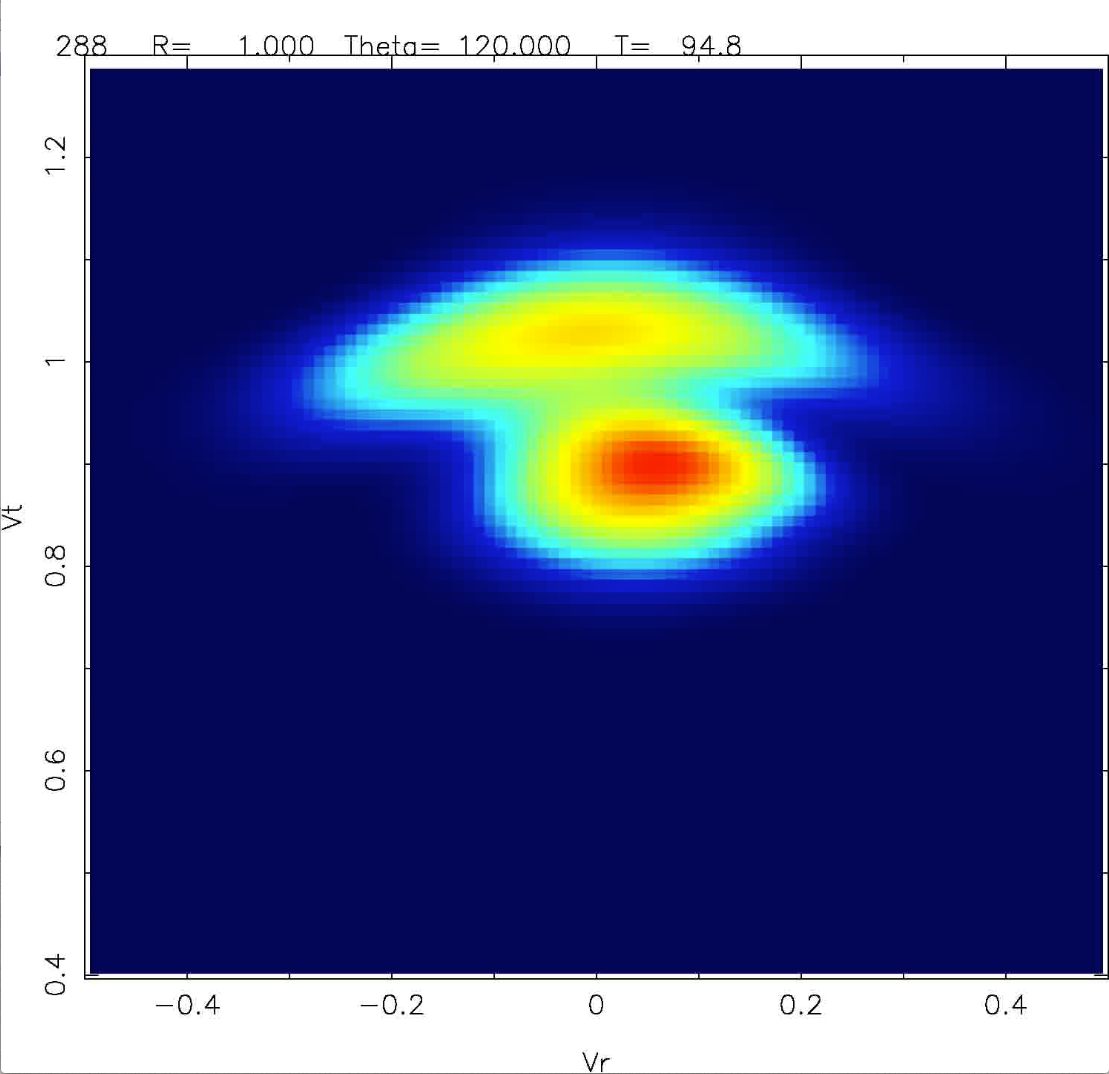}}}
\resizebox{3.9 cm}{!}{\rotatebox{0}{\includegraphics{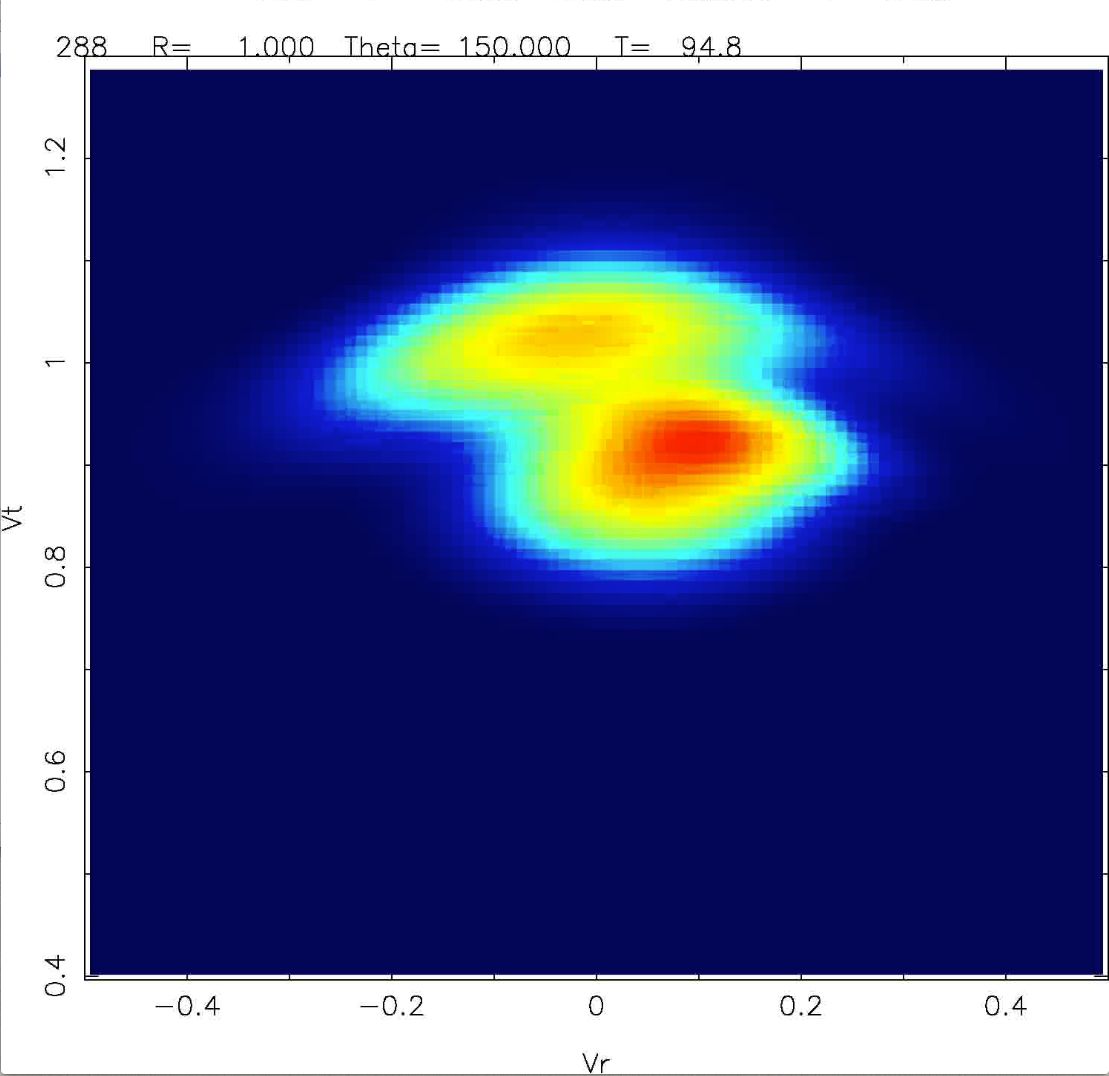}}}
\end{center}
  \caption{  ($V_R, V_\theta$) velocity distributions for model A at $R$=1.0 and $\theta=0, 30, 60, 90, 120, 150$ degrees at $T$=94.8.}    \label{fdTheta}
\end{figure*}


\begin{table}[htp]
\caption{Main characteristics of  models. }
\label{table1}
\centering
\begin{tabular}{ c c c c c c c c c c }
\hline\hline 
Model  & Grid size & $|x|$ and $|y|$ bounds & $\sigma_R$&  $\epsilon$ \\
\hline
& &  & Velocity       &  Bar  \\
& &  & dispersion   & strength \\
\hline
A       &       $288^4$ &       1.5     &0.1    &                               0.1 \\
B       &       $288^4$ &       1.5     &0.2    &                               0.1 \\
C       &       $288^4$ &       1.5     &0.1    &                               0.3 \\
D       &       $288^4$ &       1.5   &0.2      &                               0.3 \\
D'      &       $288^4$ &       2.5   &0.2      &                               0.3 \\
\hline
\end{tabular} 
\end{table}

We present four  models  (A to D, Fig \ref{densABCD} and Table 1) with two different strengths of the bar ($\epsilon=0.1$ and 0.2) and  two different values of the initial radial velocity dispersion ($\sigma_R \, =0.1$ and 0.2).  The initial velocity dispersion  is set constant with radius and it corresponds to the values of 20 and 40 km.s$^{-1}$, if $V_c=220$ km.s$^{-1}$, corresponding to the  dispersions of the young and old thin discs at the solar position, for instance  see  \cite{woj16}.

For these models,  Fig \ref{densABCD} shows the contrast of the density $\rho (x,y,t)$$-$$\rho(x,y,t$=0): the initial exponential disc density  is subtracted, making discernible  the structures within the stellar disc. We  note the two rings of overdensity at  radius  $R=0.4$ and  $R=1$,  close to the corotation and  to the OLR.       At time $T=95$ or 25 bar rotation, the orientation of the bar, rotating counterclockwise,   is   34 degrees from the $x$-axis.  In the case of the strongest bar, we also note an enhancement of the density within the rings. The enhancement is  located at the extremities of the bar in the case of the corotation ring, and perpendicularly to the bar orientation in the case of the OLR ring. Close to the solar position,  $R\sim1$ and at time $T=95$, a nearly stationary state is reached. This corresponds to  $\sim$ 25 bar rotations  or   15 Galactic rotations of stars (i.e. $\sim 3.6$\,Gy in Galactic timescale). Faint spiral-like structures are  also visible. Not surprisingly, structures seen within the disc are stronger in the case of  the strongest bar, and  fatter in the case of  the highest initial velocity dispersions.

\subsection{Resonnances}

\begin{figure}[!htbp]
\begin{center}\
\resizebox{3.9cm}{!}{\rotatebox{0}{\includegraphics{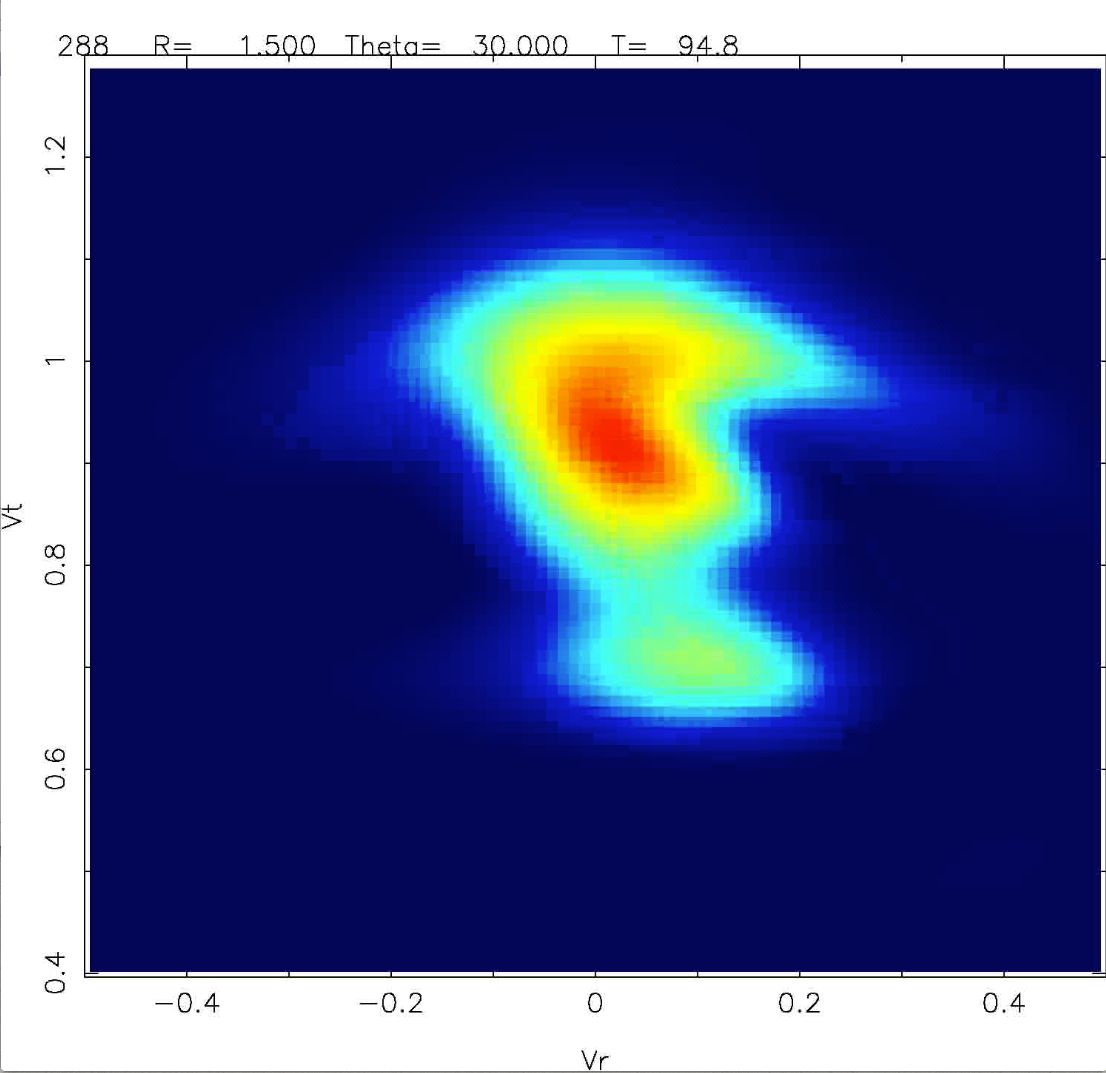}}}
\end{center}
  \caption{  ($V_R, V_\theta$) velocity distribution at $R$ = 1.5 and $T$ = 94.8. The  main orbits is the nearly circular orbit and the secondary one is a (1,1) resonant orbit.}    \label{fd11}
\end{figure}

In the vicinity of resonances, the computed DFs split in two or more components associated to the resonant orbits.  Close to the OLR, the velocity distribution $(V_R, V_\theta)$   is bimodal. The backbone of these two streams are  two (1:2) resonant orbits   aligned and anti-aligned to the bar.  Figure \ref{fdR} shows the variations of the  velocity distribution ($V_R,  V_\theta$) at different positions. It shows the relative position of the two maxima and their relative amplitude  that change along a Galactic radius  orientated at a  30 degree angle with respect to the bar main axis (approximately the observed angle between the Galatic centre - Sun axis and the bar main axis).  Figure \ref{fdTheta}   shows the changes of the same two streams  at  the OLR  ($R$ = 1) by varying the position angle with respect to the bar (0 or 180 degrees is the alignment along the main bar axis, and +30 to 45 degrees is  the approximate position of the Sun,   backwards of the direction of the rotation of the bar 
\citep{ant09,ant12,ant14,deh98,deh99,deh00,bov10,min07,min10,mon14,mon15,mon17}.
At larger radii, we also identify another important kinematic signature, the (1:1) resonance (Fig \ref{fd11}), obtained with a simulation of model-type D but with  a larger grid and x-y steps (xmax=ymax=2.5). We note that the  two velocity maxima are separated by about $\Delta V_\theta=$44 km.s$^{-1}$  and this structure should be located at $\sim$ 3 kpc from the Sun towards the Galactic anticentre. It could be detectable with existing proper motion surveys.

\subsection{Stationarity}

From the CBE in a stationary frame with a constant angular velocity \citep[][eq. 4.284]{bin08}  and a potential with two axis of symmetry, we can deduce the following symmetry relation for the distribution functions:
$f(x,y,u,v)=f(x,-y,u,-v)=f(-x,y,-u,v)$,
if the $x$ and $y$ axis are the axis of symmetry of the potential.
It implies that along the two axis of the bar $\theta_{sym}$ = 0 and $\pi/2$, then
$f(R,\phi_{sym}, v_R,v_t )=f(R,\phi_{sym}, -v_R,v_t )$.
Along these axis, the DF $f$ is even with respect to $v_R$ . For the same reasons, the final stationary density distribution must  have the same $x$ and $y$ axis of symmetry. As noted by  \citet{fux01} and \citet{muh03}, this property can be used to estimate the degree of stationarity achieved within the simulations.

\begin{figure}[!htbp]
\begin{center}\
\resizebox{4.4cm}{!}{\rotatebox{0}{\includegraphics{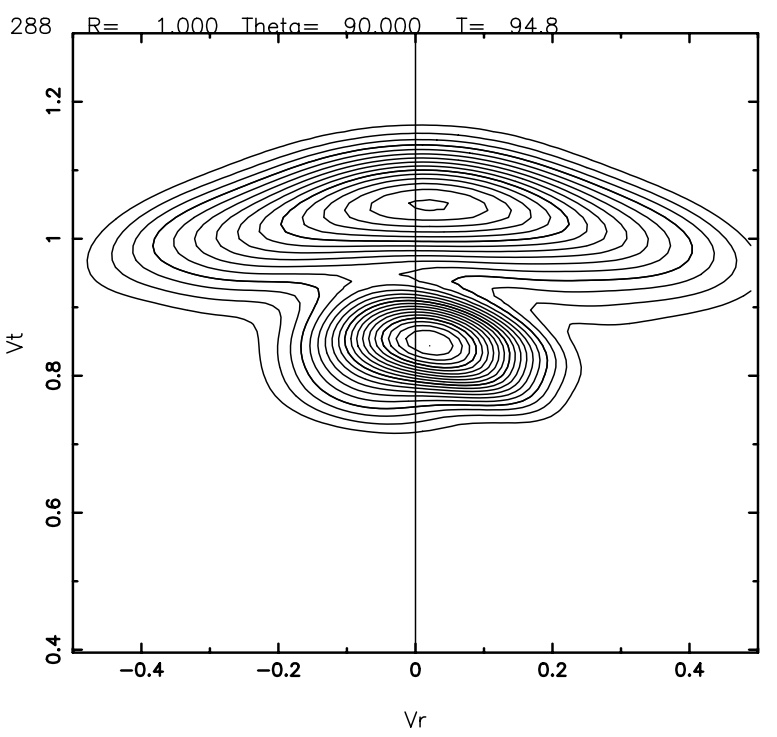}}}
\resizebox{4.4cm}{!}{\rotatebox{0}{\includegraphics{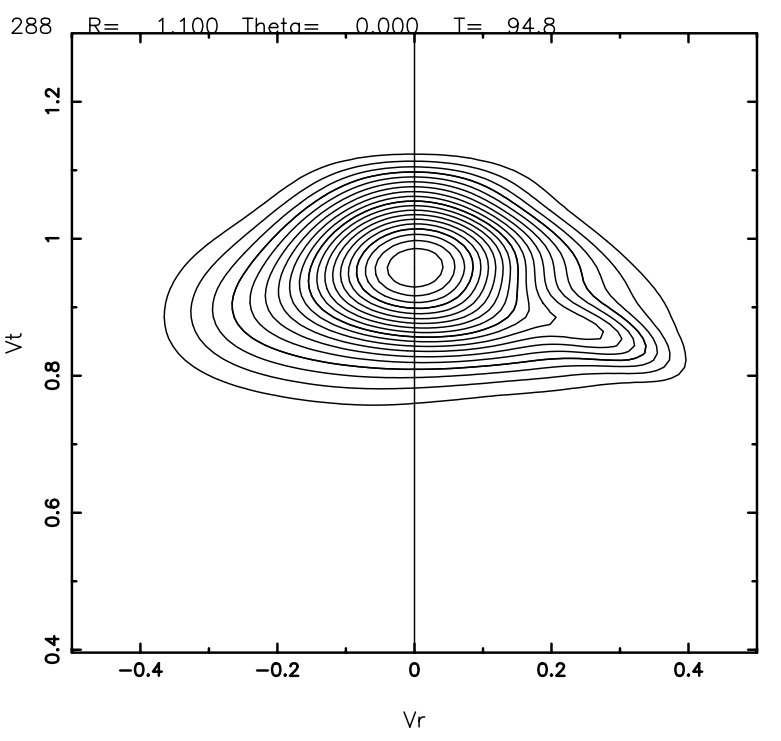}}}
\end{center}
  \caption{  ($V_R,\,V_{\theta}$) velocity distribution for model C, with a strong bar and a small velocity dispersion, at two positions. {\bf (left)} ($R$ = 1, $\theta$ = 90 degrees),  {\bf (right)} ($R$ = 1.1, $\theta$ = 0 degree). One orbital family has the $V_r$ = 0 axis symmetry, the other  one, not yet fully phase mixed has not. }
  \label{fs1234}
\end{figure}

 Figure \ref{densABCD} shows the density distribution of A to D models at $T$ = 95 where the symmetries are immediately recognizable. It is at the corotation radius that the departure from symmetry is the largest. However, in these regions and also towards the centre, the velocity distributions  are very symmetric and  smooth. They do not present any visible structure or separated streams. 

Initially, we were expecting to see different stellar streams close to the corotation, streams associated to the periodic orbits identified by \cite{ath83}. This is not the case and  looking at the exact shape of periodic orbits in our potentials at the corotation, we realise that the orbital shapes are significantly  different from these in  \cite{ath83}.  The fact that we use a simple quadrupole for the bar potential appears insufficient to correctly model  the inner part of the galaxy. For this reason, we postpone the analysis of the kinematics in the inner part of the galactic models to a future work.

At  OLR, $R=1$; we    note  that symmetry is achieved in the density profile along the major bar axis, but not along the minor bar axis. This is also visible in the velocity distribution: only at $\phi=0$ the mean radial velocity $\bar{v}_R$  is nearly null.

The $(V_ R, V_\theta)$ distribution at the OLR ($R$ = 1, $\phi$ = 90) (Fig \ref{fs1234}, left), shows two streams or orbital families.  The orbital family   with the largest $V_\theta$  has a   symmetric distribution, while the other one has a slight inclination.  Looking more precisely at this second family, we see that  the orbits  close to the periodic orbit ($V_R$ = 0, $V_\theta$ = 0.85)  are tilted in this diagram and thus are not yet phase-mixed.  However for this family, the orbits distant from the periodic orbit show a symmetry with respect to the $V_R$=0 axis and are phase-mixed.  In Fig \ref{fs1234} right, the asymmetry on the positive $V_R$ side is related to this same orbital  family. 
 
 Finally, if we examine the same distributions shown in Fig \ref{fs1234} at very different steps $T$ = 32, 64, or 94, (respectively 9, 17 and 25 bar rotation) we do not see visible evolution of the streams. This    signifies that  the phase space mixing is  much slower than the dynamical time of the galaxy and that the phase mixing will not be achieved 
 over the duration of the run performed here.
 
\begin{figure}[!htbp]
\begin{center}\
\resizebox{7.8cm}{!}{\rotatebox{0}{\includegraphics{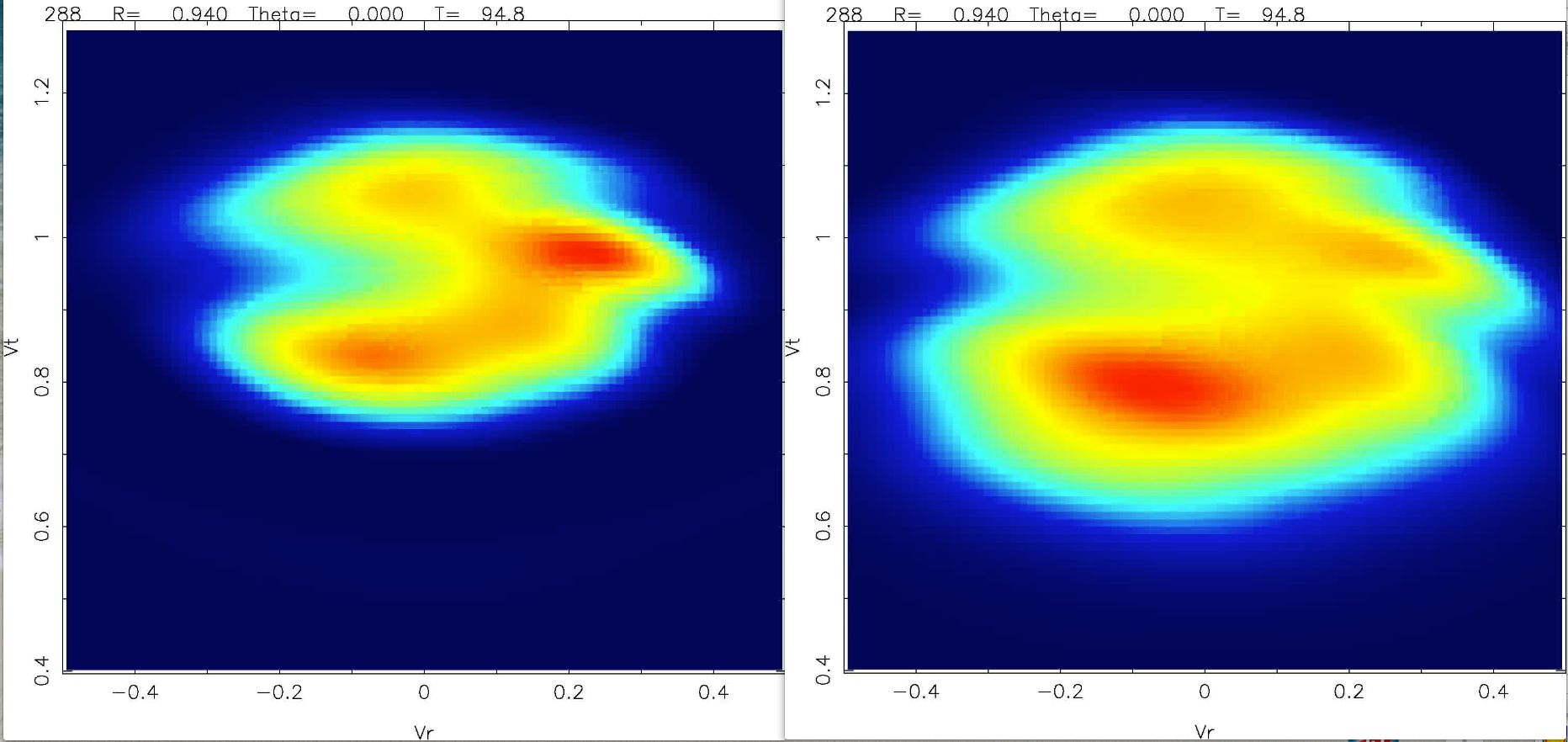}}}

  \caption{ Close  to the main axis of the bar, four streams in the case of the strong bar at $T$ = 95,  $R$ = 0.94, and $\theta$ = 0 with {\bf (left)} $\sigma_R$ = 0.1 
    {\bf (right)}~$\sigma_R$~=~0.2.
  }    \label{f5}
\end{center}
\end{figure}

Figure \ref{f5} shows an extreme case of lack of symmetry (at $R$\,= 0.94 and $\theta$ = 0). In that narrow region of the Galaxy, these  asymmetric structures  are   likely related to the initial conditions. There, the observed streams should give poor constraints on the the galactic gravitational potential.

 In Fig. \ref{fd11}, $R=1.5$, $\theta$ = $+30$ degrees, with a (1:1) orbit, the departure from symmetry (seen by comparison of the same plot at $R=1.5$, $\theta$ = $-30$ degrees) is the "plume" located at ($V_R$ = 0.2, $V_\theta$ = 1).  This reveals  the difficulties that will arise when models will be compared with Gaia observations  in order to identify the nature of observed streams within the disc.
 
\subsection{OLR streams}

Our results obtained close to the OLR can be  directly compared with   other studies   that model the Hercules stream in the solar neighbourhood 
\citep{fam05,fam08}, a stream usually explained  by  the proximity of the bar OLR  with the solar position, see  
\citet{deh98,deh99,deh00,ant09,ant12,ant14,min07,min10,bov10,mon14,mon15,mon17},
 other interpretations have been proposed : \citet{sel10,per17}
 but  see also \citet{mon17}.

The CBE modelling allows us to obtain a  smoother representation of the velocity field, by contrast with   N-body or test-particle simulations limited by the Poisson noise. The CBE resolution  gives us  a fine quantification of the stream properties  and  models accurately the faintest structures. We find many similarities  between our results from the CBE resolution and from published N-body approaches, and we  confirm  conclusions previously published about the possible location of the Sun with respect to the Galactic bar. 

Our results are qualitatively  similar to the  $(V_R,V_\theta)$  distribution at various positions  as  shown for instance by 
\citet[][Fig. 2]{deh00}, \citet[][Fig.~1]{min10}, and \citet[][Fig. 2]{bov10}. This is the reason we do not reproduce  the corresponding plots from  our own simulations.  The relative position and  amplitude of the two (1:2) streams vary rapidly depending on the Galactic position angle and  radius.  From  the comparison of models with the observations of $(U,V)$ velocities from Hipparcos measurements,  we can deduce  the position of the Sun and also the  pattern speed of the bar.

\subsection{Comparison with  solar neighbourhood observations}

Based on the analysis of stellar proper motions and distances from Hipparcos observations, \citet{deh00} identified the Hercules stream,  distinctly separated from the main stellar stream. Based on 3D velocity data, \citet{fam05} also identified a clear-cut separation. Later based on RAVE data, \citet{ant14} studied the evolution of the Hercules stream with  Galactic radii. Recently using Gaia DR1/TGAS data complemented with LAMOST data, \cite{mon17}    identified the separation between these two streams over an extended range of Galactic radii towards the Galactic anticentre

The characteristics of the stellar streams seem to depend on the different observed stellar populations. We also note that the presence of substructures  makes   a precise quantitative comparison difficult. The model-observation comparison is also limited by our poor knowledge of the distance to the Galactic centre, of the circular velocity at the solar position, and   of the solar velocity with respect to the LSR (if it were possible to  define unambiguously these two last quantities for a barred galaxy). For all these reasons, we will not discuss these points further.
 
 All these  elements limit the possibility of building an accurate quantitative model for comparison of  the separation between the observed  streams, their relative orientation and their relative amplitude. This will be certainly improved thanks to the future Gaia data by increasing the  stellar sample size  and also  the space volume probed.  Here we will consider only a rough model calibrated to recover to position and velocity of the Sun, we set $V_c$ = 220 km/s, $R$ = 8.5 kpc, and assumme unknown the LSR velocity.  In doing so, we reach  the same conclusion as \citet{ant14}: it relates the Galactic angular velocities (of the bar and of the circular orbit at the Sun Galactic radius) to the position angle of the Sun with respect to the major bar axis (positive in the bar counter rotating direction). If $\theta$ ranges from 24 to 45 degrees, they obtain a range for $\Omega_b/\Omega$  from 
1.80 to 1.90.

Close to the OLR and  from the perspective of our models, we notice that  the relative amplitude the two streams,  their respective position, and their orientation in the ($V_R, V_\theta$) plane,  quickly vary with the Galactic position  (radius $R$ and bar angle $\theta$). For a given position, increasing $\epsilon$, the strength of the bar, results in an increase of the number of stars within the Hercules stream. It also increases the separation between the two streams. Finally,  we note that  our numerical simulations show that  increasing the velocity dispersion  changes  the ratio number of stars between the two streams. It also significantly increases the separation between  the streams, this is expected from the separation of periodic orbits that depends on the bar strength, see figure 2 in \citet[]{ath83}.

We find,  as \citet{ant14} does,  that the relative aspect of the two streams  remains approximately identical over a large domain. The  Hipparcos double stream aspect (Hercules plus main stream) is seen in our models with the correct  orientation  if $R$ is within the range 1 to 1.4 (varying the  $\theta$ position). If we restrict the possible angles in the range of accepted values between 24 to 45 degrees, we obtain for the ratio  $\Omega_b/\Omega$ a range from 1.77 to 1.91 with $R$ varying respectively from 1.02 to 1.10, so just beyond the OLR. This gives us a "high" angular velocity for the bar. If $R$ = 8.5 kpc and $V_c$ = 220 km/s, then $\Omega_b$ is between  46 and 49 km.s$^{-1}$.kpc$^{-1}$.

Other  effects have not been considered in the present analysis, such as the exact shape of the circular velocity curve, a more realistic bar potential, or the modification of    scale lengths of the Galactic disc for the density $R_\rho$ and for the kinematics $R_\sigma$. However, an important effect not considered is  the non-stationarity. We have used quasi stationary models for  comparison with data. Their density is effectively nearly constant after $T$ = 30 (8 bar rotation), and then the relative aspect of the two streams close to the OLR  varies  very slowly. But, if the Galactic bar is quickly evolving, a direct comparison with our models would be partly questionable.
 This  would introduce a supplementary uncertainty for the determination of the solar neighbourhood position within the models.

This preliminary work allows us to verify the reliability of the BCE numerical resolution to analyse stellar streams, and it shows an improvement to model faint structures in the phase space by comparison with N-body simulations. Future works are planned and we will replace the quadruple galactic bar with more realistic gravitational potentials. This is necessary to analyse the behaviour of streams close to the corotation. More realistic potentials will be also necessary for a comparison to  GAIA observations. We will implement the algorithm of Filbet (2001) that minimizes numerical diffusion and oscillations in the neighbourhood of strong density and velocity gradients. Such an algorithm will improve the resolution.

From symmetry arguments, we have noticed that we did not achieved exact stationary solutions for the DFs. We suppose that this could be achieved by forcing symmetry of the distribution functions during the numerical evolution. We expect it will shorten the numerical time to reach a stationary state. This will allow a direct  comparison with the stationary solutions obtained by analytical means by  \cite{mon17b} and  with theoretical predictions.

After having considered the effect of bar resonances, we will examine the impact of  resonances of spiral arms. We intend to study the combined effects of  bar and spiral arms and the related non stationarity of the potential. This is frequently advocated as an explanation of the split of stellar streams in smaller ones as they are seen  with Hipparcos and  GAIA data. Since these structures are faint in  current N-body simulations, and are at the limit of the Poisson noise fluctuations, it will be fruitful   to reexamine this question by solving the Boltzmann equation.

\section{Conclusion}

The code presented in this paper solves the collisionless Boltzmann equation (CBE)  in a four-dimensional phase space, two-dimensional in the physical space (2D) and two-dimensional in the velocity motion (2V). It is applied to the study of the stellar kinematics within the disc of a barred galaxy.
 
We have shown that a numerical resolution of the CBE can be used to  model the stellar kinematics of a spiral galaxy. We numerically recovered  the (1:2) resonnance of the OLR created by a rotating bar that is  usually advocated to explain the  main stellar  streams observed within the solar neighbourhood. We recovered similar results  to these obtained by different authors using N-body simulations \citep{deh00,min10,ant14}. The CBE code   cancels the statistical noise allowing us to follow  faint structures and densities  within the  phase space. We  confirm  the  probable  position of the Sun with respect  to the Galactic centre and the Galactic bar orientation as well as the bar pattern speed  found by \cite{deh00} and \cite{min10}. Recent analyses of the disc stellar kinematics on  larger scales in the solar neighbourhood \citep{ant14,mon17} corroborate the interpretation of the Sun position as being close to the bar OLR. Our simulations confirms these findings.

Gaia data will provide more  accurate informations about the two (1:2) resonant orbits, and also on the Galactic bar orientation. The bar pattern speed and the position of other important resonances as the corotation   will be also constrained more tightly.  The partial phase mixing of stellar orbits will probably make  this task  laborious.  We expect that  the kinematic signatures seen in our simulation, like the  (1:1) resonant orbit  at $R$ = 1.5 (i.e. 12 kpc), should be detected. All such features  will help to constrain the disc kinematics and the Galactic gravitational potential.

\begin{acknowledgements}

We would like to thank  Ivan Minchev for a very  opportune question while we recently met at the Gaia meeting in Nice, and Benoit Famaey and Christian Boily for pertinent remarks.

\end{acknowledgements}

\bibliographystyle{aa} 
\bibliography{v2} 

\end{document}